\documentstyle[prb,aps,epsfig,float]{revtex}
\begin{document}
\flushbottom
\draft          
\wideabs{       
\title{
        Magnetotunneling in a Two-Dimensional Electron-Hole System Near Equilibrium}
\author{
        E. M. Gonz\'alez*, Y. Lin and E. E. Mendez}
\address{
        Department of Physics and Astronomy, SUNY at Stony Brook, Stony Brook, NY 11794-3800}
\date{\today}
\maketitle
\begin{abstract}
We have measured the zero-bias differential tunneling conductance of InAs/AlSb/GaSb/AlSb/InAs heterostructures at low temperatures (1.7\,K $<$ T $<$ 60\,K) and under a magnetic field at various angles with the heterostructure's interfaces. Shubnikov-de Haas oscillations in the magnetoconductance reveal the two-dimensional (2D) character of the electrons accumulated at the InAs interfaces and yield their number in each of them. The temperature dependence of the oscillations suggests the formation of a field-induced energy gap at the Fermi level, similar to that observed before in simpler 2D-2D tunneling systems. A calculation of the magnetoconductance that considers different 2D densities in the two InAs electrodes agrees with the main observations, but fails to explain features that might be related to the presence of 2D holes in the GaSb region.
\end{abstract}
\vspace{-.4cm}
}               

Heterostructures based on the InAs-AlSb-GaSb system are at the heart of low-dimensionality interband tunneling diodes. This system is also attractive from the point of view of basic physics because the unusual ordering of the InAs and GaSb energy bands allows to study the tunneling properties of combined two-dimensional electron and hole gases (2DEG and 2DHG, respectively). 

Two alternative arrangements are usually considered: GaSb/AlSb/InAs/AlSb/GaSb (type A) and InAs/AlSb/GaSb/AlSb/InAs (type B), in both of which electrons and holes accumulate at the InAs and GaSb interfaces, respectively. In type A, when the InAs layer is thin, a quasi-square InAs quantum well (QW) for electrons is created between AlSb barriers, and GaSb electrodes behave as sources of holes. In type B, a thin GaSb central layer forms a quasi-square QW for holes, whereas the InAs end layers act as electron sources. In either combination, when a voltage (V) is applied between the electrodes, negative differential resistance (NDR) appears when the bandgap of GaSb starts blocking the tunneling of electrons from the conduction band of InAs into adjacent layers.~\cite{Soderstrom:89,Luo:89}

Until now, the emphasis has been on the study of the current-voltage (I-V) characteristics, often in connection with the materials properties of the heterostructures or with the details of their band structure.~\cite{Nosho:98,Kitabayashi:98,Ogawa:00,Takamasu:93,Marquardt:96,Mendez:00} The wealth of information available on the InAs-AlSb-GaSb system under an external bias constrasts with the limited knowledge about the details of the tunneling mechanism when the system is in quasi-equilibrium, that is, in the zero-voltage limit. Since carriers accumulate in quasi-triangular wells at the interfaces, the fact that tunneling ideally occurs between quantized 2D states at the electrodes should be reflected in the conductance behavior.~\cite{Mendez:90,Eisenstein:91} Moreover, since in both types of heterostructures vertical transport involves confined valence band states (holes in GaSb) and conduction band states (electrons in InAs), one could envision tunneling taking place via a 2DH(E)G $\rightarrow$ 2DE(H)G $\rightarrow$ 2DH(E)G mechanism, with its own signature.

The 2D nature of the electronic states at the central region, whether it is InAs or GaSb, is unquestioned, but the small confinement energy at the electrodes in type A heterostructures reduces (if not eliminates altogether) the 2D character of the holes.~\cite{Mendez:92} In type B structures, however, electrons in the quasi-triangular accumulation layers at the InAs electrodes are much more confined, and in a magnetic field their cyclotron energy is large enough to yield a distinct zero-dimensionality Landau-level spectrum. These properties make InAs-GaSb-InAs based heterojunctions suitable to explore 2DEG $\rightarrow$ 2DHG $\rightarrow$ 2DEG tunneling, which is the objective of this work.

We have measured the zero-bias differential magnetoconductance (G) of InAs/AlSb/GaSb/AlSb/InAs heterostructures grown by molecular-beam epitaxy on $p^{+}$ GaSb substrates. The thickness of the undoped GaSb layer was 60 {\AA} and that of each AlSb barrier was 34 {\AA}. Each electrode consisted of an 800 {\AA} thick undoped InAs region (residual $n^{-}$ type doping estimated to be 1 to $3\times 10^{17} cm^{-3}$) adjacent to AlSb, followed by a thick (3000 {\AA} near the surface, 1 $\mu m$ next to the substrate) $n^{+}$ InAs ($N_{d}$ = $2\times 10^{18} cm^{-3}$) layer. 

Figure 1 shows the potential energy profile of the heterostructure along the growth direction, $z$. The overlap between the top of the GaSb valence band and the bottom of the InAs conduction band produces a self-limiting transfer of electrons from GaSb to InAs. As a result, an accumulation layer is formed at each InAs/AlSb interface and holes are left behind in the GaSb quantum well. Ideally the total number of holes is twice that of electrons per interface. The number of transferred electrons increases with increasing overlap energy and with decreasing barrier thickness. In the figure, $E_{0}$ indicates the ground-state energy for 2D electrons, $E_{F}$ is the Fermi level of the system, and $H_{0}$ represents the heavy-hole ground-state energy of the GaSb quantum well. The energy values have been calculated by solving Poisson and Schr\"{o}dinger equations self-consistently.

The best evidence for the presence of holes in the central well is the observation (even at room temperature) of NDR features in the I-V characteristics.~\cite{Luo:89} As the voltage between electrodes is increased, electrons from the emitter tunnel to the collector through the two barriers via the small energy window defined by $E_{F}$ and $H_{0}$. The tunneling process continues until the voltage is such that $H_{0}$ is below the emitter's $E_{0}$ level, at which point the tunnel current ceases and NDR appears. The shape of the I-V curves and the current-onset voltage depend on whether the in-plane momentum, $k_{\parallel}$, is conserved in the tunneling process. A detailed discussion of the similarities and differences between 2DEG $\rightarrow$ 2DEG and 2DEG $\rightarrow$ 2DHG is postponed for a future publication. However, it is important to mention now that, in sharp contrast with the former case, tunelling of momentum-conserving 2D electrons into 2D hole states is possible for all voltages in the range between the threshold (which depends on the difference between the number of 2D electrons and holes) and NDR voltages. 
\begin{figure}
\centerline{\epsfig{file=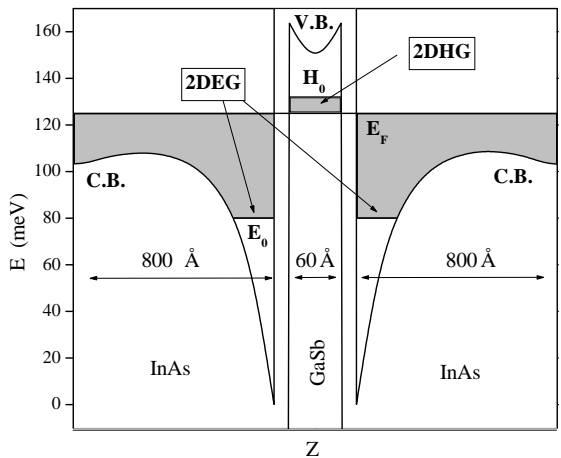,width=3in,clip=}}
\caption{Energy-band profile of a type B heterostructure in the vicinity of the Fermi level $E_{F}$. The thickness of each AlSb barrier is 34 {\AA}. $E_{0}$ and $H_{0}$ represent the confined levels for two-dimensional (2D) electrons in the InAs conduction band (C.B.) and 2D holes in the GaSb valence band (V.B.), respectively.} 
\end{figure}

We focus here on the quasi-equilibrium conductance, that is, the slope of the I-V characteristics in the V $\rightarrow$ 0 limit. The measurements were done at low temperature (1.7\,K $<$ T $<$ 60\,K) by applying a small ac modulation voltage (200 $\mu V$) between the electrodes of mesa-defined diodes with dimensions of 100 $\mu m$ $\times$ 100 $\mu m$, and a magnetic field B, up to 11\,T, forming an angle $\varphi$ with the plane of the interfaces. 

In Fig. 2(a) we show G(B) for $\varphi$ = $90^{o}$ up to B = 11\,T. The oscillations in G(B), discernible even below 0.5T, exhibit the characteristic $B^{-1}$ periodicity of Shubnikov-de Haas oscillations. They resemble those previously observed in similar measurements on type A heterostructures, which were explained in terms of the Landau levels formed in the InAs quantum well.~\cite{Mendez:93} However, there are some notable differences between the oscillations observed in both types of heterostructures. While at low magnetic fields (B $<$ 1\,T) the amplitude of the oscillations in type A increases monotonically with increasing field, that of the oscillations in Fig. 2(a) is modulated in a way similar to that observable in a two-subband system. Besides, a second set of oscillations is apparent in Fig. 2(a) at intermediate fields (1\,T $<$ B $<$ 4\,T), quite different from the spin-splitting doublets that appear at larger fields in type A. Finally, the oscillations in Fig. 2(a) have an unusual shape at high B (B $>$ 4\,T).

The temperature and angular dependence of the low-field oscillations in Fig. 2(a) have elucidated their origin. The temperature dependence of the oscillations' amplitudes (see Fig. 3) yields effective mass values between 0.025-0.028 $m_{0}$ ($m_{0}$ is the free-electron mass). The angular dependence reveals that the positions of the oscillations extrema are determined not by the total magnetic field but by its perpendicular component $B_{\perp}$, thus proving the 2D character of the charges involved in magnetotunneling. The low-field amplitude modulation and the doublets at intermediate fields indicate two different Landau-level ladders, each of them with its own $B^{-1}$ periodicity and associated 2D carrier density. From the periods of the two oscillations, a simple estimate yields $N_{1}$ = $5.13\times 10^{11} cm^{-2}$ and $N_{2}$ = $5.07\times 10^{11} cm^{-2}$. 
\begin{figure}
\centerline{\psfig{file=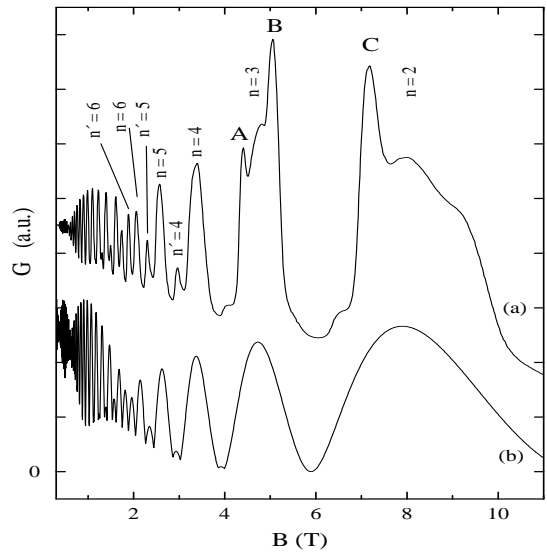,height=3in,width=3in}}
\caption{Differential tunneling conductance in the zero-voltage limit (V=0) {it vs} a magnetic field perpendicular to the layers, for the heterostructure to which Fig.1. refers. (a) Experimental curve measured at T = 1.7\,K. (b) Calculated curve using Eq. (1) in the text. The indices of the Landau levels in the two InAs electrodes are indicated as n and n'.}
\end{figure}

These results are consistent with the magneto-conductance oscillations being due to tunneling between Landau levels of the 2D electron gases formed at the InAs electrodes. In this picture $N_{1}$ and $N_{2}$ represent the 2D electron densities of those electrodes; the small difference between the two indicates that the two interfaces are almost identical and, therefore, that the subband energies, $E_{01}$ and $E_{02}$, are almost the same. Instead, it would be possible to interpret $N_{1}$ and $N_{2}$ as the carrier densities of two electron subbands of the same electrode, let us say the emitter. But if this were the case, the two densities should be quite different; indeed, our self-consistent calculations indicate that a second subband would be only marginally occupied.

The above picture---2D electrons tunneling from occupied Landau levels in the emitter to equivalent empty states in the collector---is supported by the agreement found between the experimental results of Fig. 2(a) and the calculated tunneling magnetoconductance shown in Fig. 2(b). The calculation is similar to that by Lyo~\cite{Lyo:98} for the conductance between two 2D electron gases in a perpendicular field $B_{\perp}$, 
\begin{equation}
G_{zz}(B_{\perp}) \propto g(B_{\perp})\sum_{n_{1},n_{2}} \theta_{n_{1},n_{2}}(u)^{2}\rho_{n_{1}}(\mu)\rho_{n_{2}}(\mu - \Delta),
\end{equation}
where $g(B_{\perp})$ = $eB_{\perp}/\pi\hbar$ is the Landau-level degeneracy; $\theta_{n_{1},n_{2}}(u)$ is an ``overlap" function between Landau levels $n_{1}$ and $n_{2}$ of the electrodes, which, if the Landau-level index is conserved, is 1 when $n_{1}$ = $n_{2}$ and zero otherwise; $\rho_{n_{i}}$ is the density of states (DOS) for the Landau level of index $n_{i}$; and $\Delta$ is the difference between the subband energies of the two electrodes ($\Delta$ = $|E_{01}-E_{02}|$). We have used the approximation that only Landau levels at the Fermi energy $\mu$ contribute significantly to $G_{zz}(B_{\perp})$, thus reducing the double sum on $n_{1}$ and $n_{2}$ to $n_{1}$ = $n_{1F}$ and $n_{2}$ = $n_{2F}$. We have also assumed that the value of the overlap function is 1 because since the carrier densities in the two InAs layers are almost the same, in practice, at the Fermi energy the indices of the Landau levels of the two electrodes are the same. For the DOS we have used a Gaussian distribution, 
\begin{equation}
\rho_{n}(\mu) = {1 \over {\sqrt{2\pi }\Gamma}} exp\left[ -{{(\mu -\varepsilon_{n})^{2}} \over {2\Gamma^{2}}}\right],
\end{equation}
where $\varepsilon_{n}$ = $\hbar\omega_{c}(n+1/2)$ is the energy of the n Landau level and $\Gamma$ is the level damping (or broadening), given by $\Gamma$ = $0.5\eta \hbar \omega_{c}/B_{\perp}^{1/2}$ ($\eta$ is a sample-dependent parameter). The Fermi energy, which depends on magnetic field, was determined from the condition that the total number of carriers, $2N_{0}$ = $N_{1}$ + $N_{2}$, be constant: 
\begin{equation}
2N_{0} = \int_{0}^{\infty}D^{2D}(E)f(E,\mu ,T)dE,
\end{equation}
where $f(E, \mu, T)$ is the Fermi distribution function, and $D^{2D}(E)$ is the total density of states, which includes the DOS of each accumulation layer. Since the experimental results used for comparison were obtained at 1.7\,K, in calculating $\mu$ from Eq. (3) we assumed that T = 0\,K.

As seen in Fig. 2(b), this simple model reproduces well the peak positions throughout the entire field range. Our intention has not been to find systematically a ``best" set of parameters but rather to illustrate how a set of reasonable values can explain the observed magnetoconductance oscillations. The following set has been used in the calculation of Fig. 2(b): 0.027 $m_{0}$ for the electron effective mass, $\eta$ = 0.8 $T^{1/2}$ for the parameter related with the broadening $\Gamma$, $N_{0}$ = $5.7\times 10^{11} cm^{-2}$ and $\Delta$ = 4 meV for the missalignement between the two energy subbands, $E_{01}$ and $E_{02}$. The value used for the effective mass is consistent with that determined from the temperature dependence of the low-field Shubnikov-de Haas oscillations. The subband energy difference $\Delta$ is less than 10\% of any of the two energies; in other words, the difference between $N_{1}$ and $N_{2}$ is quite small when compared to the average density in each electrode. The values of this difference and of the sum ($2N_{0}$), although somewhat larger than those estimated from the Shubnikov-de Haas periods, were chosen so that the calculated G(B) ``fitted" well the experimental G(B) for the entire magnetic-field range. The small discrepancy may be due to our assumption that $N_{0}$ is independent of B, an assumption that is especially questionable at high fields, given the physical origin of the 2D electrons (the transfer from the GaSb central well).
\begin{figure}
\centerline{\psfig{file=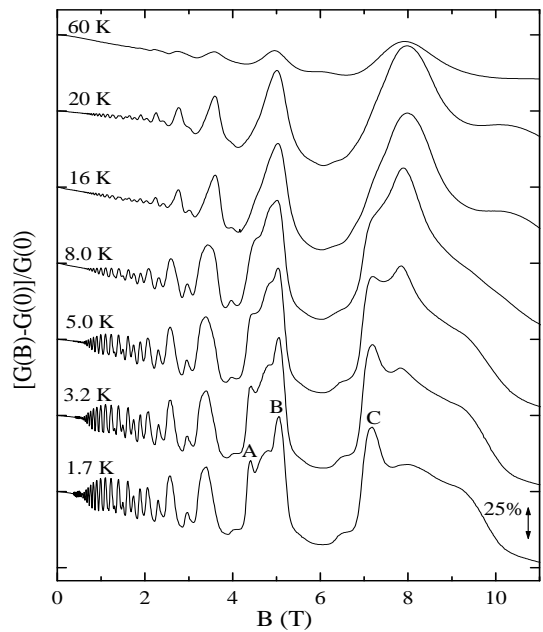,height=3.5in,width=3in}}
\caption{Temperature dependence of the magnetoconductance, relative to its zero-field value, for the heterostructure shown in Fig. 1. For clarity, the origin is different for each curve. While the amplitude of the low-field (below 3T) oscillations decreases monotonically with increasing T, that of the high-field (above 4T) broader oscillations first increases and then decreases. The features labeled as A, B and C, which are more distinguishable at the lowest T, cannot be explained by the simple model described in the text and may be related to the 2D holes in the GaSb central well.}
\end{figure}

The different carrier densities of the two (nominally identical) accumulation layers suggest an asymmetry in the two AlSb barriers, a fact confirmed by the asymmetry observed between the two polarities of the diodes' I-V characteristics. If one barrier were thicker or higher than the other, then the intrinsic transfer of electrons from GaSb to each InAs interface would be different. In practice, barrier asymmetry can happen for a number of reasons; for instance, Ga segregation during the epitaxial growth of the heterostructures would lead to an Ga-rich AlGaSb layer, which has a smaller bandgap than AlSb and would present a smaller tunneling barrier. This segregation process is known to be the more significant the higher the substrate temperature.~\cite{Nosho:98} Since our heterostructures were prepared at 500 $^{o}C$---a relatively high temperature for these materials---it would not be surprising that the barrier asymmetry is caused by that mechanism.

In spite of the overall good agreement between the experiment and the calculation, at high magnetic fields the measured conductance behaves in a way for which the model cannot account. In Fig. 2(a) there are three peaks (marked by ``A", ``B", and ``C") that do not have a counterpart in Fig. 2(b). Although the fact that they were clearly resolved only at the lowest temperatures (see Fig. 3) might suggest spin splitting effects, the inclusion of spin in the calculation (not shown in Fig. 2(b)) failed to explain both the number of peaks and the fields at which they occur. 

The temperature dependence of the broader Shubnikov -de Haas oscillations underlying peaks A, B, and C also warrants a comment. As T increases, the amplitude of the oscillations increases a bit before it decreases at high temperatures (see Fig. 3), suggesting a thermally activated behavior followed by the usual T-induced weakening of quantum effects, when $k_{B}T$ becomes comparable to the cyclotron energy and the scattering time is comparable to the cyclotron time. Although the low-temperature range investigated here is too limited to draw any conclusion, it is quite possible that the activated behavior is a consequence of the fomation of a field-induced energy gap at the Fermi level. Such a gap has been proposed to explain a similar temperature dependence of the 2DEG $\rightarrow$ 2DEG conductance in GaAs-GaAlAs heterostructures.~\cite{Eisenstein:92} Calculations have shown that the gap appears as a result of the additional energy cost required to extract an electron from the emitter Fermi sea (creating a ``vacancy") and to inject it into the collector one (creating an ``interstitial") when many-body effects (for example, Coulomb interactions) become sizeable.~\cite{Yang:93}

The quasi-equilibrium magnetoconductance experiments presented here show unambiguously the participation of 2D electrons in the 2DEG $\rightarrow$ 2DHG $\rightarrow$ 2DEG tunneling process, but do not reveal the role of the 2D holes in it. Whether the unaccounted low-temperature peaks (A, B and C) are evidence of the holes we do not know. However, it is worth pointing that, based on their shape and the temperature range in which they are most visible, those peaks resemble the features found in the in-plane magnetoresistance of InAs-GaSb heterostructures in which 2D electrons and holes coexist.~\cite{Mendez:85} Although these features remain unexplained, their connection with 2D holes is well established, since they disappear when the 2D holes are eliminated by the application of hydrostatic pressure.~\cite{Beerens:87} It is then reasonable to imagine that the extra peaks in Fig. 3 are somehow related to the quantized holes in the GaSb quantum well, but more work, both theoretical and experimental, is needed to set this possible correspondence on a firmer ground.
\acknowledgments
We thank Dr. A. Sacedon for her participation in the early stages of this work, Dr.W. I. Wang for contributing the heterostructures used in it, and Dr. Aleiner for useful discussions. The support of the US Army Research Office and of a NATO postdoctoral fellowship (to E. M. G.) have made this work possible.

\end{document}